\documentclass[pdflatex,sn-basic]{sn-jnl}

\jyear{2021}%

\theoremstyle{thmstyleone}%
\theoremstyle{thmstyletwo}%

\theoremstyle{thmstylethree}%

\usepackage{float}

\usepackage{color}
\usepackage{colortbl} 
\newcommand{\eg}{e.g., }
\newcommand{\ie}{i.e., }

\usepackage{amsmath,algorithm,algpseudocode}
\usepackage{xcolor}
\usepackage{breakcites}
\usepackage[bottom]{footmisc}
\usepackage{wrapfig}

\usepackage{url,lineno,microtype}
\usepackage[onehalfspacing]{setspace}

\begin{document}

\title[Supervised Neuronal Parameter Estimation]{Supervised Parameter Estimation of Neuron Populations from Multiple Firing Events}


\author*[1]{\fnm{Long} \sur{Le}}\email{vlongle@seas.upenn.edu}
\equalcont{Work done while student at University of Massachusetts Amherst.}

\author[2]{\fnm{Yao} \sur{Li}}\email{yaoli@math.umass.edu}

\affil*[1]{\orgdiv{GRASP Lab}, \orgname{University of Pennsylvania}, \orgaddress{\city{Philadelphia}, \state{Pennsylvania}, \country{United States}}}

\affil[2]{\orgdiv{Department of Mathematics and Statistics}, \orgname{University of Massachusetts Amherst}, \orgaddress{\city{Amherst}, \state{Massachusetts}, \country{United States}}}


\abstract{The firing dynamics of biological neurons in mathematical models is often determined by the model's parameters, representing the neurons' underlying properties. The parameter estimation problem seeks to recover those parameters of a single neuron or a neuron population from their responses to external stimuli and interactions between themselves. Most common methods for tackling this problem in the literature use some mechanistic models in conjunction with either a simulation-based or solution-based optimization scheme. In this paper, we study an automatic approach of learning the parameters of neuron populations from a training set consisting of pairs of spiking series and parameter labels via supervised learning. Unlike previous work, this automatic learning does not require additional simulations at inference time nor expert knowledge in deriving an analytical solution or in constructing some approximate models. We simulate many neuronal populations with different parameter settings using a stochastic neuron model. Using that data, we train a variety of supervised machine learning models, including convolutional and deep neural networks, random forest, and support vector regression. We then compare their performance against classical approaches including a genetic search, Bayesian sequential estimation, and a random walk approximate model. The supervised models almost always outperform the classical methods in parameter estimation and spike reconstruction errors, and computation expense. Convolutional neural network, in particular, is the best among all models across all metrics. The supervised models can also generalize to out-of-distribution data to a certain extent.}




\maketitle

\section{Introduction}\label{sec:intro}

It is well known that neurons in our brain can produce very
complicated spiking patterns, including a few different types of
neural oscillations. Some oscillations can be reproduced by
mathematical neuron models at a certain level. In particular, a
spiking pattern named multiple firing event (MFE) is observed in many
neuronal network models \cite{Chariker,rangan2013dynamics,rangan2013emergent,zhang2014coarse}. In an MFE, a certain proportion
of neurons in the population, but not all of them, fires a spike during
a relatively short time window and forms a spike volley. Similar spiking
patterns that lie between homogeneity and synchrony have been observed
in many experimental studies. It is believed that MFEs are responsible
for the Gamma rhythm in the central nervous system \cite{rangan2013dynamics,rangan2013emergent, henrie2005}. In general, MFE
can be observed in neuronal populations with both excitatory and
inhibitory neurons when the parameters are suitable. An MFE
is caused by a balance of recurrent excitation and inhibition from neurons. Spikes of excitatory neurons excite both
excitatory and inhibitory populations. The former induces a cascade of
spiking activities, while the latter forms an inhibitory current that
stops the spiking volley. Neuronal networks with MFEs are intrinsically multi-scale because of the rapid spiking activities during MFEs. 

Despite the intuitive mechanism of MFEs and some early investigation about the low dimension nature of MFEs \cite{cai2022model}, it is very difficult to find a low dimensional dynamical system to accurately approximate the MFEs. Known
results about MFE mechanism in \cite{zhang2014coarse,li2019stochastic, cai2022model} do not provide a full
answer to that. Spiking activities in MFEs can range from quite
homogeneous to very synchronous. However, no existing theory can accurately predict the spiking
pattern without running the full model, nor infer parameters from the
spiking activities. In this paper, we attempt to shed some light on
this challenging problem by training supervised machine learning models to learn the
spiking activities of MFEs. More precisely, we generate many spike series from a wide range of parameters using a stochastic neuronal network model introduced in \cite{Li2017HowWD}. These series are then labeled by
the corresponding parameters as a training set. After some training, supervised models including a convolutional neural network, deep neural network, random forest, and support vector regression can backwardly infer the parameters from which the spike series were obtained. 

The result is very encouraging. Although the neuronal population model
has a lot of stochasticity and the MFEs have high volatility, our supervised models, especially the convolutional neural network (CNN), successfully grasp the key relation between parameters
and spiking patterns. When an input spiking pattern is given to the CNN, the predicted parameters can produce a
visually similar spiking pattern. These similarities can be quantified by various reconstruction error measures. The resulting reconstruction errors from supervised models are low and can be largely attributed to the inherent stochasticity of the neuronal network model that generated the data. Further, the supervised models are benchmarked against traditional approaches in parameter estimation including a genetic search, Sequential Neural Posterior Estimation (SNPE), and a random walk approximate model.

Further numerical experiments also confirm that supervised models, especially neural networks, have some generalization
ability. When the spike series is generated by a parameter set that is deliberately sampled outside of the training set, the supervised models can still reconstruct the spike series reasonably well. In both the test and generalization experiments, the convolutional neural network has the best performance while incurring a relatively low computation cost.

The organization of this paper is as follows. Section \ref{sec:related} reviews some classical approaches in the literature and their drawbacks. Section \ref{sec:meth} describes the supervised learning approach, the data-generating neuronal network model, data generation process, a regularization heuristic to aid learning, and some classical models that we benchmarked against. In Section \ref{sec:res}, we report the results on the test and generalization datasets. Section \ref{sec:con} is the conclusion. Appendices \ref{sec:pars} and \ref{sec:model_configs} detail the parameter settings and model configurations used.

\section{Related Work} \label{sec:related}

\begin{figure}
\begin{center}
\hspace*{-1cm}
    \includegraphics[scale=0.15]{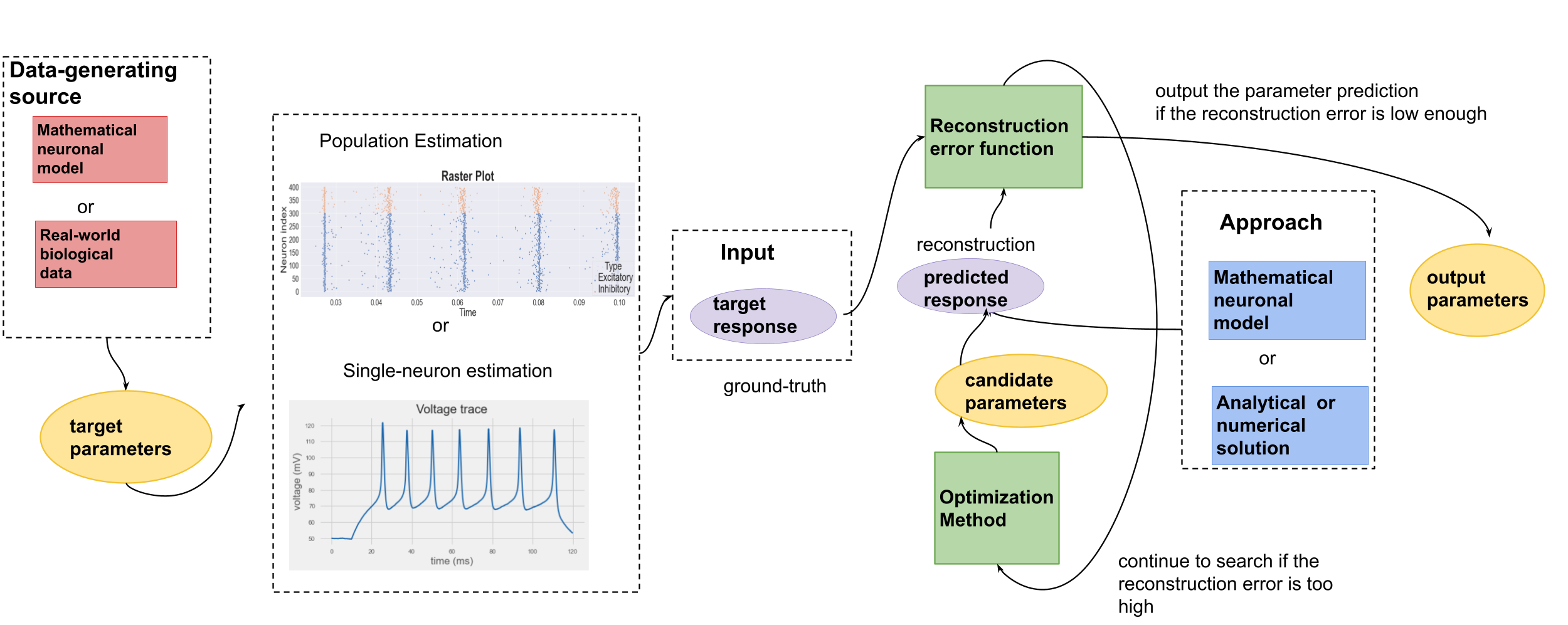}
\end{center}
\caption{An overview of traditional approaches to parameter estimation. A target response is given. An optimization routine is used to produce a candidate parameter set. The parameter set is then used to produce a predicted response either by a simulator or some approximation solution. The predicted response is compared against the target one using some reconstruction error function. This procedure ceases when the predicted response is deemed similar enough to the target.}
\label{fig:related}
\end{figure}

Parameter estimation is a difficult and well-studied problem in the literature. Most traditional approaches proceed as follows (see Figure \ref{fig:related}). A target spike count input, in the case of population estimation, or a voltage trace, in the case of single-neuron estimation, is given, either from synthetic data generated by a mathematical model or from biological experiments (\eg \cite{MCMC, evol}). The input is typically vectorized, for example by discretizing into time
series or extracting summary features such as the average spike height and firing rate \cite{review}. Then, an optimization framework is used to search through the parameter space and select a candidate parameter set that is likely to produce a neural response similar to the target response. The reconstructed response can be obtained in two main ways. Some methods use a neuronal model as a simulator. Others obtain the signal through solving some differential equations, describing the dynamics of the neurons. For example, \cite{pde} solves a Fokker-Planck PDE numerically while \cite{single} obtains the solution to the Hodgkin-Huxley neuronal model \cite{hodgkin} analytically. Sometimes, an approximate model where an analytical solution is available, for example
a Poisson process in \cite{pde}, or some other surrogate models \cite{zhang2020DNN} are used. The reconstructed response is then compared to the ground truth using some loss function. If the loss is low enough, then the current parameter set is outputted, and this procedure stops. Otherwise, we continue to search through the parameter space.

There are many search methods available such as likelihood-free Bayesian inference \cite{bayesian, snpe}, evolutionary algorithm \cite{evol}, Simplex \cite{simplex}, interior point line search \cite{offset}, particle swarm optimization and MCMC \cite{MCMC}, and interval analysis based optimization \cite{single}. The paper \cite{review} reviews some other search algorithms such as simulated annealing and gradient descent. 

In those works, there is additional time required in running extra simulations (simulator-based) or in numerically solving some differential equations (solution-based) to construct predicted responses at inference time. The computation cost in iteratively searching through the parameter space can also be high. In the case of solution-based approaches, expert knowledge is also required in constructing the differential equations, and their analytical solutions or approximation. Table \ref{tab:approaches} compares our approaches to some others. Note that the supervised learning approach in this paper does not require additional simulations other than the ones used to
generate the training data or any expert knowledge. Most works that we encountered (including \cite{evol, MCMC, snpe}) estimate the parameters of a single neuron. In our work, we found that parameter estimation at a population level using these methods requires simulating a large number of neuron populations at inference time and therefore sharply increases the computation expense.

\begin{table}[ht]
\centering
\caption{Comparing some approaches to parameter estimation.}
\begin{tabular}{ | c | c | c | } 
  \hline
  \rowcolor[gray]{.9} Approach & Require simulations at inference? & Solution-based \\ 
  \hline
   \cite{evol}                   & yes & no \\
  \hline
  \cite{bayesian}                & yes & no \\
  \hline
     \cite{simplex}              & yes & no \\
  \hline
    \cite{pde}                   & no  & yes \\
  \hline
     \cite{offset}               & yes & no \\
  \hline
    \cite{MCMC}                  & no  & yes \\
        \hline
 \cite{single}                   & no  & yes \\
  \hline
     \cite{lfp}                  & yes & no \\
        \hline
      Supervised learning (ours) & no  & no \\
        \hline

\end{tabular}
\label{tab:approaches}
\end{table}

The supervised learning approach here is most similar to the work of \cite{lfp}, which uses a convolution neural network to learn neuronal parameters from local field potentials (LFPs). The key difference between our work and theirs is that while they extract high-level features, namely 6-channels LFPs, from spiking trains as input, we use the raw spike series. Obtaining local field potentials requires replaying the generated spike trains
from a neuron model to a more biophysically detailed model, thus adding more computation cost. Deep learning models have been observed to be able to automatically extract high-level and useful features from data (for example, in image processing \cite{cnn_review}, natural language processing \cite{text} or neuron response features \cite{snpe}). Thus, we chose to feed in the raw spike series as inputs. 

The paper \cite{zhang2020DNN} uses a deep neural network (DNN) to learn a surrogate forward model mapping neuronal parameters to high-level features of the spike trains, \ie their firing rates. Then, another outer search is required to do parameter tuning similar to Figure \ref{fig:related}. In contrast to this approach, we use high-dimensional spike series as inputs, which allow us to learn the inverse mapping from neuronal responses to parameters directly. 

Another work \cite{snpe} also uses neural networks. There, neural networks were used as density estimators in a sequential Bayesian framework, also requiring a lot of simulations at inference time.

The supervised learning approach explained in the next section will make a distinction between training and inference stages. Most computation time is invested in the training stage so that inferences on a new data point can be done almost instantaneously.

\section{Methods} \label{sec:meth}

\begin{figure}
\begin{center}
    \includegraphics[scale=0.3]{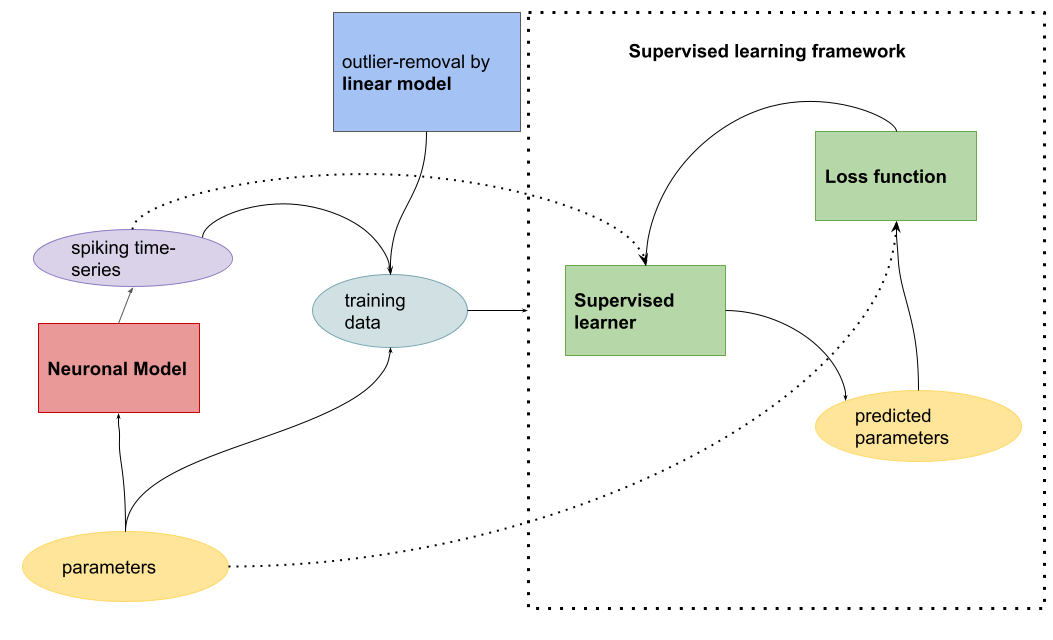}
\end{center}
\caption{Supervised learning framework to parameter estimation. The training data consisting of the spiking time series and parameter labels is fed to a supervised learning model. To ensure the quality of the training data, we also use an outlier removal procedure (see Section \ref{sec:outlier}).}
\label{fig:ours}
\end{figure}

\subsection{Supervised Learning Approach} \label{supervised}
The approach we chose is to frame the parameter estimation as a supervised learning problem (see Figure \ref{fig:ours}). Note that, unlike previous search approaches that we have discussed, supervised learning requires the parameter labels as inputs along with the usual spiking time series. We will discuss more the process of constructing spiking time series in Section \ref{sec:data}. Requiring parameter labels is usually not a severe constraint since there is often some mechanistic model available to generate synthetic data. For those mechanistic models, we can control the parameters, and thus know their values.

This supervised learning approach proceeds by generating a large number of parameter set and spike series pairs. At training time, supervised models have to learn to perform accurate parameter prediction. In general, a supervised learner tries to minimize the average prediction loss on the training set (\ie the empirical risk)
\begin{equation}
 \mathcal{L} = \frac{1}{N} \sum_{i=1}^N L(\underline{y}_i, \underline{\hat{y}_i})
\end{equation}
where $N$ is the training set size, $\underline{y}_i$ is the true parameter set, $\underline{\hat{y}_i}$ is the estimated parameter set, and $L$ is some loss function. In this paper, we choose $L$ to be the mean absolute error (MAE) \ie
$L(\underline{y}, \underline{\hat{y}}) = \ell_1(\underline{y} - \underline{\hat{y}})$ where $\ell_1$ is the L1 norm. MAE was chosen since we have found that it is less sensitive to outliers than other loss functions such as the mean squared error.

At inference, a target spike series is given without a label, and the models will then try to recover the parameters without needing additional simulations. We also include an outlier removal
procedure Section \ref{sec:outlier} to ensure the quality of the training data.

\subsection{Neuronal Model} \label{sec:neuronal_model}
In this section, we will describe the neuronal model used to generate data. Mathematical neuronal models in the literature come in great variety, varying in complexity. For example, the complex Hodgkin-Huxley equations \cite{hodgkin} model each ion channel within a neuron explicitly while much simpler mean-field methods such as the Wilson-Cowan equation \cite{meanfield} models some averaged quantities of neurons such as firing rate over time. In choosing which neuronal model to use, there is a trade-off between biological realism and tractability: more realistic models tend to contain quantities that are harder to measure experimentally or computationally expensive to simulate. A popular class of models at an intermediate level of complexity is integrate-and-fire \cite{iaf1, iaf2}, which can generate a diverse set of firing dynamics. The model that we use for this study, from \cite{Li2017HowWD},  is of the integrate-and-fire class with some known theoretical properties. 

\subsubsection{Description of the Model} \label{neuronal_model}
The model used in this paper is a stochastic integrate-and-fire type. Instead of modeling physiological details such as ion channels,
we only keep track of the electrical property of a neuron --
its membrane potential. The membrane potential of a neuron changes
after receiving external or in-network stimuli. A spike is fired when the membrane
potential reaches a certain threshold potential. 

\begin{center}
\begin{figure}[H]
    \centering
    \includegraphics[scale=0.5]{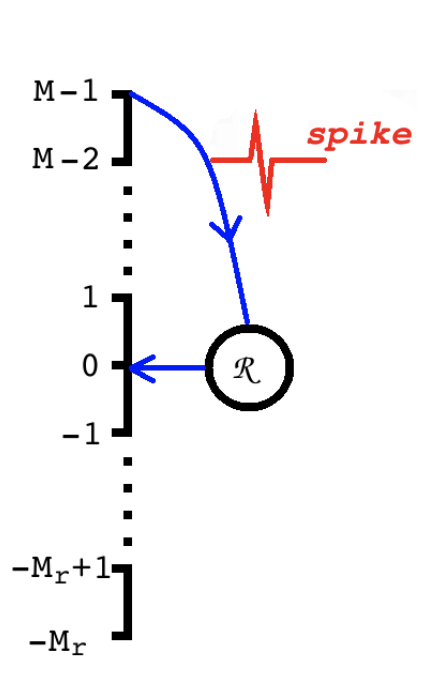}
    \caption{Each neuron's membrane potential is discretized into the range $[-M_r, M] \cup \mathcal{R}$. The voltage is set to $\mathcal{R}$ after a spike. When the neuron comes out of refractory, the voltage starts at $0$.}
    \label{fig:spike_time}
\end{figure}
\end{center}

The model consists of a local population of $N_E$ excitatory (E) and $N_I$ inhibitory (I) neurons. Each neuron has a membrane potential, which we assume to take on finitely many values in $\{-M_r, -M_r + 1, ..., 0, ..., M\} \cup \{\mathcal{R}\}$, where $M_r, M \in \mathbb{N}$ and $\mathcal{R}$ is a special state, called refractory. A neuron in this refractory state is ``asleep" and cannot be affected by stimulus. The minimum possible potential $-M_r$ is known as the reversal potential. When the voltage $V_i$ of the $i^{th}$ neuron reaches the voltage threshold $M$, the neuron is said to
\textit{spike} or \textit{fire} and
$V_i$ is set to $\mathcal{R}$. In other words, the neuron enters a refractory period after spiking (see Figure \ref{fig:spike_time}).

There are two mechanisms for changing the voltage $V_i$: background stimuli and
neuron-to-neuron interactions. In the background, there is an external
drive that increases the membrane potentials. This external drive can
be thought of as inputs from a neighboring neuronal population or
from sensory input. Mathematically, we represent the external inputs as Poisson processes delivering impulsive kicks to each neuron independently. These inputs always increase the membrane
potentials by $1$. We have two Poisson arrival processes parametrized by $\lambda^E, \lambda^I > 0$, representing the rates of the Poisson kicks to $E$ and $I$ neurons respectively. Under these Poisson processes, the time between two consecutive kicks is exponentially distributed, and the number of kicks over a given time interval is Poisson distributed.

Within the population, there are opportunities for changing membrane potentials whenever a neuron spikes. When a neuron fires, the neuron sends a signal (via neurotransmitters) to its post-synaptic neurons. The set of post-synaptic neurons is randomly and dynamically chosen as needed. Specifically with $Q', Q \in \{E, I\}$, when a Q'-type neuron fires, each Q-type neuron has the probability of $P_{QQ'}$, independent of other neurons, of being connected to the spiking neuron. $P_{QQ'} \in [0, 1]$
is called the connectivity probability. As such, the set of post-synaptic neurons to a given neuron is not fixed and is chosen anew every time. This follows the convention from the paper \cite{Li2017HowWD}. 

Each signal from $Q'$-type neuron to $Q$-type neuron carries a weight of $S_{QQ'}$, where $S_{QQ'} \in \mathbb{Z^+}$ if $Q'=E$ and $S_{QQ'} \in \mathbb{Z^-}$ if $Q'=I$. That is to say, an excitatory signal increases the post-synaptic neuron's voltage while an inhibitory signal decreases the voltage. Upon arrival to the post-synaptic neuron $i$, the signal alters $V_i$ precisely by $S_{QQ'}$ \ie \ $V_i =
V_i + S_{QQ'}$. There is a random delay in neuron-to-neuron signal
delivery. Each signal arrives at its destination after an
exponentially distributed delay time with mean $\tau^{Q'}$. Note that the delay time of each signal is only dependent on the type of presynaptic neuron (excitatory or inhibitory) and is independent between signals.

\begin{center}
\begin{figure}[H]
    \centering
    \includegraphics[scale=0.5]{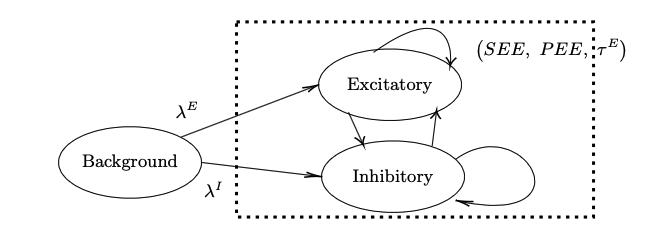}
\caption{There are two types of stimuli: background and neuron-to-neuron. The background inputs are parametrized by $\lambda^E, \lambda^I$, depending on the type of receiving neurons. The in-population interaction from $Q'$-type to $Q$-type neuron is parametrized by $(S_{QQ'}, P_{QQ'}, \tau^{Q'})$ -- the kick strength, connectivity probability and mean delay time.}
\label{fig:pop}
\end{figure}
\end{center}

We now describe a neuron's refractory period. After a neuron spikes,
it enters a recovery period known as refractory. During this period,
any background kick or signal from other neurons will not alter the
neuron's potential. The neuron stays in refractory for an
exponentially distributed amount of time with mean
$\tau_{\mathcal{R}}$. After this period, the voltage is set to $0$ and the
neuron behaves as usual with respect to stimuli. 

The illustration of the neuronal model's description is in Figure \ref{fig:pop}. Table \ref{tab:params} summarizes all of the parameters in this neuronal model.

\begin{table}[ht]
\centering
\caption{A list of parameters in the data-generating neuronal model.}
\begin{tabular}{ | c | c | } 
  \hline
  \rowcolor[gray]{.9} Parameters & Description \\ 
  \hline
   $M$             & Maximal neuronal voltage (firing threshold) \\
     \hline
    $-M_r$         & Minimum neuronal voltage \\
  \hline
   $N_E$           & Number of excitatory neurons in the population. \\
  \hline
   $N_I$           & Number of inhibitory neurons in the population. \\
  \hline
    $\lambda_E$    & Rate of external stimulus to E-population. \\
  \hline
      $\lambda_I$  & Rate of external stimulus to I-population. \\
  \hline
     $P_{EE}$      & Probability of E-to-E dynamic synaptic connection. \\
  \hline
    $P_{IE}$       & Probability of E-to-I dynamic synaptic connection. \\
  \hline
    $P_{EI}$       & Probability of I-to-E dynamic synaptic connection.  \\
  \hline
      $P_{II}$     & Probability of I-to-I dynamic synaptic connection. \\
  \hline
        $S_{EE}$   & Synaptic strength for E-to-E connection. \\
  \hline
          $S_{IE}$ & Synaptic strength for E-to-I connection.  \\
  \hline
          $S_{EI}$ & Synaptic strength for I-to-E connection. \\
  \hline
          $S_{II}$ & Synaptic strength for I-to-I connection. \\
  \hline
      $\tau_R$     & Mean refractory time \\
  \hline
        $\tau_E$   & Mean E-kick delay time from an excitatory presynaptic neuron \\
  \hline
        $\tau_I$   & Mean I-kick delay time from an inhibitory presynaptic neuron \\
  \hline
\end{tabular}
\label{tab:params}
\end{table}

\subsubsection{Known Theoretical Properties}
It is proven in the paper \cite{Li2017HowWD} that the neuronal model is a
countable state Markov process that admits a unique invariant
probability distribution. In addition, the speed of convergence
towards this invariant probability distribution is exponentially
fast. In \cite{li2019stochastic,li2020entropy}, it is further shown that many statistical
properties of MFEs, including spiking count, variance, and entropy,
are both well-defined and computable under this model.

\subsubsection{Firing Dynamics}
The neuronal model can cover a wide range of spiking patterns from
almost homogeneous to fully synchronous. In this section, we demonstrate a gallery of neuronal responses produced by the model by varying its parameters. We generate three neuron populations by fixing the parameter values as given in Table \ref{tab:static_params} and means of the ranges for other parameters in Table \ref{tab:dynamic_params} while varying one parameter $\tau_E$ (mean excitatory kick delay time) as follows.

\begin{enumerate}
    \item The \textbf{``Homogeneous"} population, abbreviated as ``Hom" in the Figure. 
        \[
            \tau_E = 9 \textrm{ ms} .
        \]    
    \item The \textbf{``Regular"} population, abbreviated as ``Reg" in the Figure.
        
        \[
            \tau_E = 5 \textrm{ ms} .
        \]    
        
    \item The \textbf{``Synchronized"} population, abbreviated as ``Sync" in the Figure. 
        \[
            \tau_E = 1 \textrm{ ms} .
        \]    
       \end{enumerate}
       
\begin{figure}
    \hspace*{-2cm}
   \includegraphics[width=1.2\linewidth]{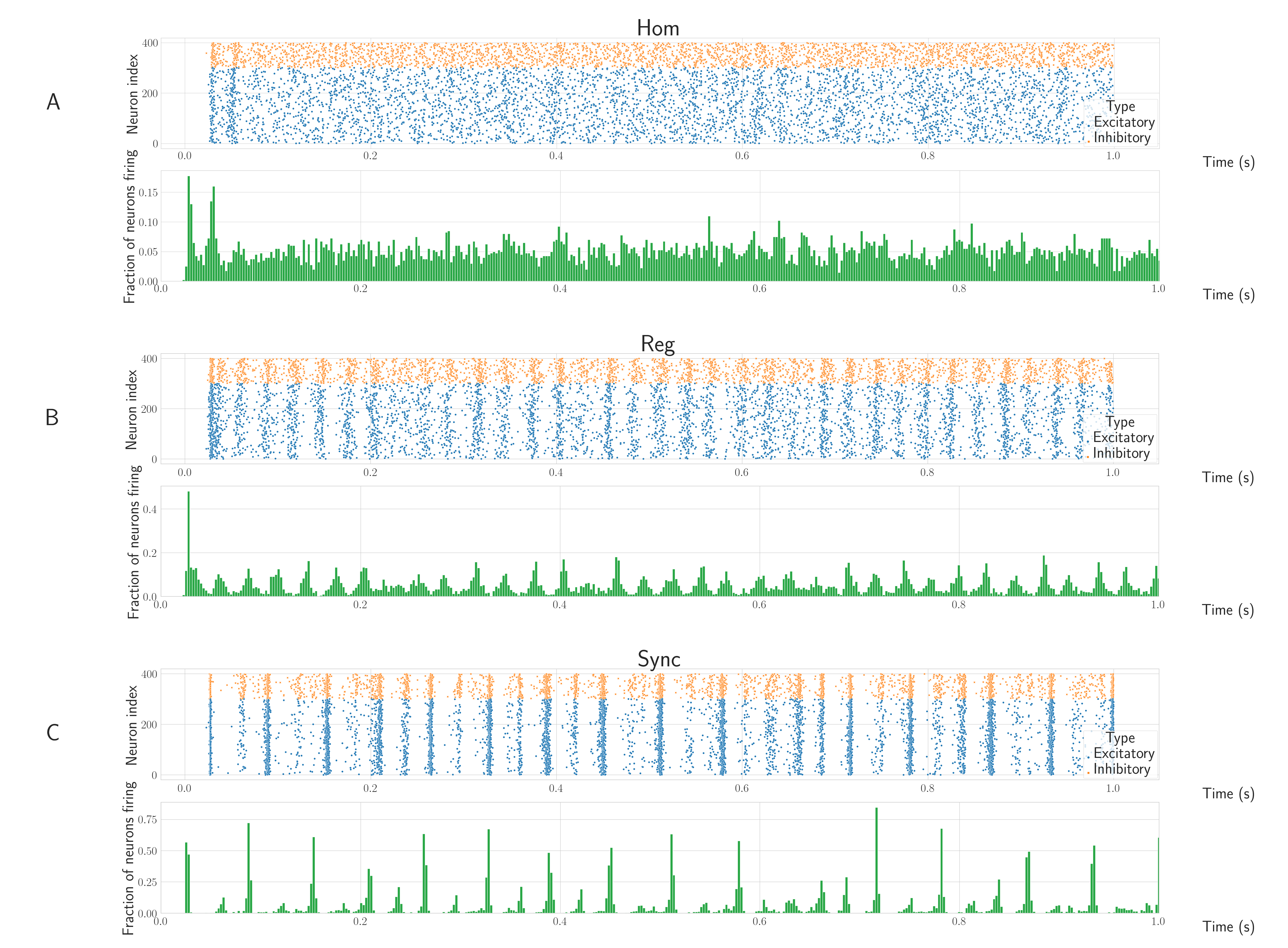}
   \caption{Effect of $\tau_E$ on neuron response. Neuronal responses of three neuronal populations, homogeneous (\textbf{Hom}), regular (\textbf{Reg}), and synchronized (\textbf{Sync}), are included. The raster plots display each firing event. Plots of the proportion of neurons in the populations firing at any given time are also included. The homogeneous population tends to have dispersive firing patterns while neurons in the synchronized population tend to fire together. The regular population displays an intermediate firing pattern between homogeneity and synchrony.}
    \label{fig:hom_reg_sync} 
\end{figure}

The neuronal response is given in Figure \ref{fig:hom_reg_sync}. As can be observed, the population with a lower $\tau_E$ displays a higher degree of synchrony. The synchronization is marked by a high number of spikes in a short duration followed by a period of low spiking activities. In the raster plots, higher synchrony is revealed by the presence of concentrated columns, indicating periods when a lot of neurons fire at once. We also include plots of the fraction of neurons in a population that are firing at a given time. In those plots, synchrony is shown by sharp peaks in the spiking fraction. In the Reg population that lies between the two extremes, one can observe some local peaks in the spiking fraction plot that look distinctive from the more uniform pattern of the Hom population's plot but not as sharp as those of the Sync population. For the Reg population, one can also see that concentrated columns are beginning to form in the raster plot. 

\begin{figure}
    \hspace*{-2cm}
   \includegraphics[width=1.3\linewidth]{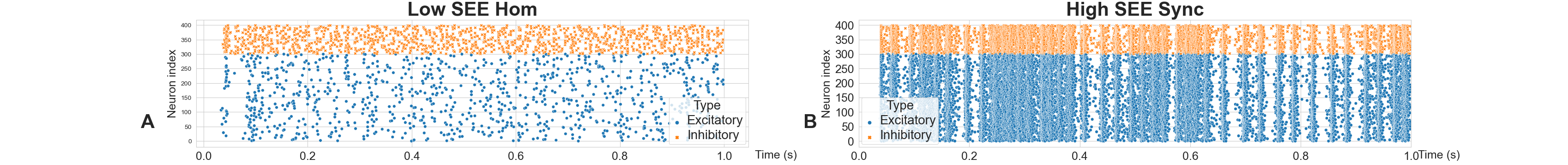}
\caption{Effect of $S_{EE}$: two neuronal populations of low and high $S_{EE}$ values. The population with the higher $S_{EE}$ value is more synchronized and fires far more frequently.}
\label{fig:SEE_raster}
\end{figure}

In Figure \ref{fig:SEE_raster}, we demonstrate the effect of varying $S_{EE}$ (E-E kick strength) on the population. We fix the population's configuration as in the
Regular population of Figure \ref{fig:hom_reg_sync}. $S_{EE}$ is varied as follows.
\begin{enumerate}
    \item The \textbf{``Low SEE Homogeneous"} population, abbreviated as ``Low SEE Hom" in the Figure. 
        \[
            S_{EE} = 2.5 .
        \]    
        
    \item The \textbf{``High SEE Synchronized"} population, abbreviated as ``High SEE Syn" in the Figure. 
        \[
            S_{EE} = 10.
        \]    
\end{enumerate}

With $S_{EE}=2.5$, it takes an excitatory neuron in the ``Low SEE Homogeneous" population $40$ consecutive excitatory kicks to spike. With $S_{EE}=10$, an excitatory neuron in the ``High SEE Synchronized" population only requires $10$ consecutive excitatory kicks to spike. Thus, we observe that the population of the higher $S_{EE}$ value has a higher degree of synchrony and spiking rate.

\subsection{Data Generation} \label{sec:data}
In this study, we consider a parameter estimation task on 6 parameters while fixing the other 11 parameters. The 11 static parameters include those specifying the size of the
neuronal population, the membrane potential's range, the external
drive strengths, the connectivity probabilities, and the mean refractory time. These parameters are given in Table \ref{tab:static_params} in the appendix. The strengths of the external drive are fixed here because they are easy to measure by traditional methods. One can
just measure the average increasing rate of membrane potential when
there is no MFE. The rest of the parameter values follow closely the setting in the paper \cite{Li2017HowWD} and are designed to be as biologically realistic as possible. 

The $6$ dynamic parameters of neuron-to-neuron kick strengths and mean delay times ($S_{EE}, S_{IE}, S_{EI}, S_{II}, \tau_E, \tau_I$) are varied to generate a diverse collection of neuronal populations. We select the delay times $\tau_E, \tau_I$ uniformly over some ranges. The kick strengths, $S_{EE}, S_{IE}, S_{EI}, S_{II}$, are first chosen uniformly over some ranges and then filtered out by the outlier removal procedure described in Section \ref{sec:outlier}. The parameter ranges are given in Table \ref{tab:dynamic_params} in the appendix.

The spiking time series is generated as follows. Given a parameter set, we instantiate a local neuronal population where each neuron has zero initial membrane potential. The population is evolved by random parametrized processes described previously in Section \ref{neuronal_model}. We monitor the population's dynamics for $2$ seconds. In each 5ms interval, we record the number of spikes from the E and I neurons separately. This process produces two spiking time histograms, one for the
E-(sub)population and the other for the I-(sub)population. The reasoning for counting I and E firings separately is that I and E neurons are very different in mechanism and how they affect the rest of the population (through excitation and inhibition), thus separate counting would produce more informative input for supervised training. We then concatenate these two histograms to produce one final series for the entire population. 

We remark that here we only use a spiking time histogram instead of the whole raster plot as the input (\eg by feeding raster plots as images into supervised models) because neurons in this model are interchangeable. When a neuron fires, its post-synaptic neurons are decided on-the-fly. However, it is known that the MFE dynamics of neuronal network model with randomly generated static connection graph is essentially similar \cite{li2020entropy}. Hence we expect our method remain to hold for a large class of neuronal network models with static graph structures.

\subsection{Regularization Heuristic} \label{sec:outlier}

Many sampled parameter sets will be biologically unrealistic, producing extreme responses. For example, consider two extreme populations in Figure \ref{fig:extreme}. The parameters along with the neuron firing rates
$f_E, f_I$ of those populations are given below. In both populations, $\lambda_E =
\lambda_I = 3000$ and $\tau_E = \tau_I = 2$ ms. All of the other parameters are fixed from values from the last Section.    

\begin{enumerate}
    \item The \textbf{``Extreme Excitation"} population.
        \[
            S_{EE} = 7.0, \quad S_{IE} = 1.0, \quad S_{EI} = -0.5, \quad S_{II} = -4.0.
        \]    
        \[
            f_E = 129.29, \quad f_I = 71.64
        \]
        
    \item The \textbf{``Extreme Inhibition"} population.
        \[
            S_{EE} = 3.0, \quad S_{IE} = 5.0, \quad S_{EI} = -4.0, \quad S_{II} = -0.5.
        \]    
        \[
            f_E = 0.02, \quad f_I = 22.93
        \]
\end{enumerate}
The resulting raster plots given in Figure \ref{fig:extreme} are not biologically realistic. Thus, identifying those bad parameter sets can accelerate supervised learning. For this, we use an approximate scheme introduced in \cite{Li2017HowWD}. The scheme is a differential equation to approximate the firing rate of the neuronal network. For a $Q$-type neuron, the membrane potential $v_Q$ is modeled by the following equation.

\begin{equation}
    \frac{dv_Q}{dt} = F^{+}_Q - F^{-}_Q,
\end{equation}
where $F^{+}_Q$ is the upward driving force consisting of background stimuli and excitatory kicks
\begin{equation}
    F^{+}_Q = f_E \times N_E P_{QE} S_{QE} + \lambda^{Q},
\end{equation}
and $F^{-}_Q$ is the downward force consisting of inhibitory kicks. 
\begin{equation}
    F^{-}_Q = f_I \times N_I P_{QI} S_{QI}.
\end{equation}
$f_E$, $f_I$ are the firing rates of the $E$ and $I$ populations respectively. This model assumes that $v_Q \in [0, 1]$ so we scaled the voltage range appropriately whenever this model is used. We seek a self-consistent pair $(\tilde{f_E}, \tilde{f_I})$ such that when these rates are used as parameters in the differential equation, they produce the same firing rates. It can be shown that the self-consistent pair is unique. The unique self-consistent pair $(\tilde{f_E}, \tilde{f_I})$ is given below.

\begin{equation}\label{eqn:linear1}
    \tilde{f_E} = \frac{\lambda_E(M+C_{II}) - \lambda_I C_{EI}}{(M-C_{EE})(M+C_{II})+(C_{EI}C_{IE})}
\end{equation}
\begin{equation}\label{eqn:linear2}
    \tilde{f_I} = \frac{\lambda_I(M+C_{EE}) - \lambda_E C_{IE}}{(M-C_{EE})(M+C_{II})+(C_{EI}C_{IE})}
\end{equation}
where $C_{QQ'} = N_{Q'} P_{QQ'} S_{QQ'}$ if $Q'=E$ and $C_{QQ'} = -N_{Q'} P_{QQ'} S_{QQ'}$ if $Q'=I$. Note that the paper \cite{Li2017HowWD} uses the convention that $Q_{IE}, Q_{II}$ are non-negative while those parameters are negative in the current study so we take care of that by reversing the signs of $C_{QQ'}$ when needed. Further, note that this model does not take the delay times $\tau_E, \tau_I$ into account, and is only a simple approximation to the stochastic neuron model. However, it serves as an useful heuristic to identify unrealistic
parameter sets. 

\begin{figure}
    \hspace*{-2cm}
   \includegraphics[width=1.3\linewidth]{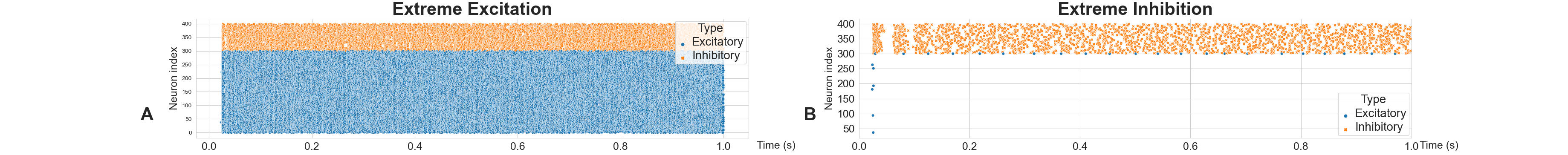}
\caption{The raster plots of two extreme parameters. These two parameter sets were removed from our training dataset by the outlier removal procedure.}
\label{fig:extreme}
\end{figure}

The approximation model breaks down for the two bad examples from Figure \ref{fig:extreme}. The self-consistent firing rates for the ``Low SEE Hom" network are $\tilde{f_E} = -13.51, \tilde{f_I} = 3.73$ and for the ``High SEE Syn" network are $\tilde{f_E} = -1.64, \tilde{f_I} = 14.71$. It does not make physical sense for a firing rate to be negative. Thus, such extreme parameters can be flagged as questionable. One way to utilize this heuristic is to use it as a regularization term of the loss
function in the training procedure of neural networks. In other words, we can augment the regular loss function $L$ with a regularization term to produce the augmented loss $L_A$.
\begin{equation}
    L_A(\underline{y}, \underline{\hat{y}}) = L(\underline{y}, \underline{\hat{y}}) + \lambda g(\underline{\hat{y}})
\end{equation}
where $\underline{y}$ is the target parameter set and $\underline{\hat{y}}$ is the candidate parameter set, $\lambda \in \mathbf{R}$ is the regularization coefficient (a hyper-parameter), and $g$ is a function that determines the biological feasibility of $\underline{\hat{y}}$. Candidate parameters that produce self-consistent firing rates that are negative or too large are deemed infeasible. We have
\begin{equation}
    g(\underline{\hat{y}}) = g_E({\underline{\hat{y}}}) + g_I({\underline{\hat{y}}})
\end{equation}
where
\begin{equation}
    g_Q(\underline{\hat{y}}) = \max(0, \tilde{f_Q}(\underline{\hat{y}}) - H) + \max(0, - \tilde{f_Q}(\underline{\hat{y}}) - H)
\end{equation}
where $Q \in \{E, I\}$, $\tilde{f_Q}(\underline{\hat{y}})$ is the self-consistent firing rate parametrized by $\underline{\hat{y}}$ from equation \ref{eqn:linear1} and \ref{eqn:linear2}, and $H$ is a high value ($200$ in our experiment). This $g_Q$ function penalizes those $\hat{\underline{y}}$ values that lead to $\tilde{f_Q}$'s falling outside the realistic range $[0, H]$.

However, in our pilot experiment, we found that the $g$ term destabilizes training, leading to non-convergence of the loss. It is also not applicable to supervised models that do not have an explicit loss function. Therefore, we chose to use the self-consistent approximation as an outlier removal procedure instead. Whenever a parameter set is generated as a candidate for the training dataset, we use the approximation to check if the self-consistent firing rates $\tilde{f}_E, \tilde{f}_Q$ are both in the range $[0, H]$, and only include the parameters if that is the case. 

\subsection{Classical Models} \label{sec:classical}
We compare the performance of the supervised models against that of three classical models: genetic search, Sequential Neural Posterior Estimation (SNPE), and a random walk model. The genetic algorithm (\cite{genetic}) was used for parameter estimation in the paper \cite{evol}. SNPE was used for parameter estimation in papers \cite{bayesian, snpe}. The random walk model is developed in \cite{Li2017HowWD} as an approximation to the data-generating neuronal model. For all of these algorithms, a
single target spike series (without the parameter label) is given at a time. The algorithms then search through the parameter space to find the parameter set that best reconstructs the target series.

The genetic algorithm starts out with some sample parameter sets. It then evolves these sets through mutation and a fitness function. The fitness function we used is the inverse of the mean absolute reconstruction error \ie 
\begin{equation}
\textrm{fitness}(\underline{\hat{y}}) = 1/\ell_1(\underline{x} - x(\hat{\underline{y}}))
\end{equation}
where $\hat{y}$ is the candidate parameter set, $\ell_1$ is the L1 norm, $\underline{x}$ is the target spike series, and $x(\hat{\underline{y}})$ is the predicted spike series, constructed by running the neuronal model simulator (Section \ref{sec:neuronal_model}) using the candidate parameter set.

The SNPE algorithm works by iteratively updating the estimated posterior distribution of the parameter set. SNPE starts with some initial parameter sets sampled from a prior distribution. The parameter sets are used to generate spike series using the simulator. The reconstructed series are compared against the target series, and the posterior distribution can be refined. This method also uses a neural network as a conditional density estimator $q(\underline{y} \mid \underline{x})$ that
models the distribution of the parameter set $\underline{y}$ given the spike series $\underline{x}$. 

The hyper-parameters for the genetic and SNPE algorithms are given in the Appendix \ref{sec:model_configs}.

Comparing to simulation-based approaches, the random walk method is solution-based. It estimates the parameter set $(S_{EE}, S_{IE}, S_{EI}, S_{II})$ while fixing other parameters. The random walk method (\cite{Li2017HowWD}) models the membrane potential of a neuron as a random walk driven by excitatory, inhibitory and external inputs. The three sources of inputs are assumed to be independent Poisson processes. Then, it was shown that the membrane potential of a type Q neuron is an irreducible Markov jump process that admits a unique stationary distribution
$\underline{\nu_Q}$ that depends on some given firing rates $f^{in}_E, f^{in}_I$ fed into the model. $\nu_Q$ can be computed by solving the following system of linear equations 
\begin{equation}
    \label{eqn:linear_sys}
    \begin{cases}
        A_Q \underline{\nu_Q} = \underline{0} \\
       \underline{1}^T \underline{\nu_Q} = 1 
        \end{cases}
\end{equation}
where $A_Q$ is the generator matrix of the Markov jump process, and \underline{1} is the all-one vector.

Given the stationary distribution $\underline{\nu_Q}$, the out-of-the-model firing rates $f_E^{out}, f_I^{out}$ can be defined as follows
\begin{equation}
    \label{eqn:out_rates}
    f_Q^{out} = N_E P_{QE} f^{in}_E \sum_{i=M-S_{QE}}^{M-1} \underline{v}_Q(i) + \lambda^Q \underline{\nu_Q}(M-1).
\end{equation}
A self-consistent pair $(\tilde{f_E}, \tilde{f_I})$ is then a pair of rates such that when it is used as the input into the random walk model, the pair of out-of-the-model rates that comes out is itself \ie $\tilde{f_E} = f_E^{in} = f_E^{out}, \tilde{f_I} = f_I^{in} = f_I^{out}$. The paper \cite{Li2017HowWD} shows that fixing the neuronal population parameters, there exist such a self-consistent firing rate pair $\tilde{f_E}, \tilde{f_I}$. Further, these rates were found to be unique in their numerical
simulations. Thus, in this approximation scheme, we use a trust-region optimizer to numerically search for a self-consistent pair for a given $S_{QQ'}$ set candidate. The idea of using some numerical iteration methods to find self-consistent solutions or parameters is not new, for example see the Schrodinger-Poisson equation solver in Physics \cite{physics}. 
Once we have found a self-consistent $(\tilde{f_E}, \tilde{f_I})$, we can also approximate the self-consistent interspike variance $\tilde{\sigma_E}, \tilde{\sigma_I}$. The interspike variance $\tilde{\sigma}_Q$ is the variance of the distribution of the elapsed time between two consecutive spikes of a single neuron from the subpopulation Q. 

Let $Y_t^Q$ denote the membrane potential of a Q-neuron and $G^Q(Y_t, dt)$ the change in potential of over a small period $dt$. The interspike random walk model in the paper \cite{Li2017HowWD} assumes that
\begin{equation}
    \label{eqn:brownian_motion}
    Y^Q_{t + dt} = Y^Q_t + G^Q(Y_t, dt)
\end{equation}
\begin{equation}
    G^Q(Y_t, dt) \approx S_{QE} \textrm{Pois}(N_E \tilde{f}_E P_{QE}dt) - S_{QI} \textrm{Pois}(N_I \tilde{f}_I P_{QI}dt) + \textrm{Pois}(\lambda^Q dt).
\end{equation}
where Pois($\alpha$) denotes a Poisson distribution with rate $\alpha$. It is well known that a standard Poisson process $N_t$ can be approximated by $t + B_t$, where $B_t$ is a Wiener process. Hence $Y^Q_t$ can be approximated by a stochastic differential equation
\begin{equation}
    \label{eqn:SDE}
    \mathrm{d} Z^Q_t = \tilde{f}_Q \mathrm{d}t + \hat{\beta}_Q \mathrm{d}B_t \,,
\end{equation}
where 
$$
\hat{\beta}_Q = \sqrt{S_{QE}^2 N_E \tilde{f}_E P_{QE} + S_{QI}^2 N_I \tilde{f}_I P_{QI} + \lambda^Q}.
$$
Then the first arrival time of $Z^Q_t$ from $0$ to the threshold $M$ is given by an inverse Gaussian distribution $IG(\tilde{f}_Q^{-1}, M^-2\hat{\beta}_Q^{-2})$. The self-consistent interspike variance $\tilde{\sigma}_Q = \tilde{f}_Q^{-3}\hat{\beta}_Q^2 M^2$ is the variance of this inverse Gaussian distribution. Therefore, the network parameters $(S_{EE}, S_{IE}, S_{EI}, S_{II})$ to the statistics $(\tilde{f}_E, \tilde{f}_I, \tilde{\sigma}_E, \tilde{\sigma}_I)$ is a mapping from $\mathbb{R}^4$ to $\mathbb{R}^4$. The numerical inverse of this mapping can be used to find parameters. 

For a given parameter set, we compute the self-consistent statistics, \ie firing rates and interspike variances, through a trust-region optimization. Then, we can compute the reconstruction error between the self-consistent statistics and the target statistics. An outer trust-region optimizer is used to guide the search over the parameter space to minimize that reconstruction error. This procedure is given in Algorithm \ref{alg:random_walk}. The output of Algorithm \ref{alg:random_walk} is the predicted (best candidate) parameter set $(\hat{S}^*_{EE}, \hat{S}^*_{IE}, \hat{S}^*_{EI}, \hat{S}^*_{II})$ that maps to the corresponding self-consistent statistics $(\tilde{f}^*_E, \tilde{f}^*_I, \tilde{\sigma}^*_E, \tilde{\sigma}^*_I)$, which are most similar to the target statistics $(f_E, f_I, \sigma_E, \sigma_I)$.

The purpose of the approximate model is to avoid having to run the full neuronal model simulator, which is computationally expensive. However, the drawback is that this mean-field scheme assumes the arriving spikes are timely homogeneous. Since MFEs are spike volleys that occur in a short time frame, the random walk approximation of both firing rate and interspike interval has some error, especially when the MFEs are very synchronous. In addition, since the MFE dynamics usually concentrates at the vicinity of a low dimensional set, the mapping from parameter space to the target statistics is likely to be singular. Hence the inverse mapping may have large derivatives, which further amplifies the error of the random walk approximation. 

Due to computational cost, we ran the random walk on 1000 examples and ran SPNEC and genetic search on 100 examples. Note that unlike machine learning models that are trained on different examples, these three models by design do not benefit from being run on more examples so these runs are primarily for model evaluation.

\begin{algorithm}
\caption{Random walk approximation scheme. $\underline{S}$ is the target statistics. The algorithm has an outer and an inner trust-region optimizer. The outer trust-region optimizer search through the space of candidate parameter sets $\underline{\hat{y}}$ to minimize the reconstruction error between the reconstructed and target statistics. Given a candidate $\underline{\hat{y}}$, the inner trust-region computes the self-consistent statistics $\underline{\hat{S}}$ to minimize the consistency error using the random walk approximation.}
\label{alg:random_walk}
\begin{algorithmic}[1]
\Function{inverseMap}{$\underline{S}$} \label{alg:a}

    \Return \Call{trustRegionOptimizer}{\textsc{reconstructionError}}.solution \Comment{\parbox[t]{.2\linewidth}{Outer optimization over the candidate parameter sets $\underline{\hat{y}}$}}
\EndFunction
\Statex

\Function{reconstructionError}{$\underline{S}$, $\underline{\hat{y}}$} \label{alg:b}
    \State $\underline{\hat{S}} \gets $ \Call{forwardMap}{$\underline{\hat{y}}$} \Comment{Compute the self-consistent statistics of the given $\underline{\hat{y}}$}
    \State \Return $||\underline{\hat{S}} - \underline{S}||_2$
\EndFunction
\Statex

\Function{forwardMap}{$\underline{\hat{y}}$} \label{alg:c}
    \State $\tilde{f}_E, \tilde{f}_I \gets $ \Call{trustRegionOptimizer}{\textsc{ConsistencyError}} \Comment{\parbox[t]{.3\linewidth}{inner optimization over the space of rates $f_E^{in}, f_I^{in}$ to obtain a self-consistent solution.}}
    \State $\tilde{\sigma}_E, \tilde{\sigma}_I \gets H(\tilde{f}_E, \tilde{f}_I)$ \Comment{\parbox[t]{0.5\linewidth}{Computed using the first passage time of the approximate Brownian motion in equation \ref{eqn:brownian_motion}}}
    \State $\underline{\tilde{S}} \gets [\tilde{f}_E, \tilde{f}_I, \tilde{\sigma}_E, \tilde{\sigma}_I]$

    \Return $\underline{\tilde{S}}$
\EndFunction
\Statex

\Function{ConsistencyError}{$\underline{\hat{y}}$, $\underline{S}^{in}$}
\State $f_E^{out}, f_I^{out} \gets$ \Call{computeOutStats}{$\underline{\hat{y}}$, $\underline{S}^{in}$}
\State \Return $(f_E^{out} - f_E^{in})^2 + (f_I^{out} - f_I^{in})^2$
\EndFunction
\Statex

\Function{computeOutStats}{$\underline{\hat{y}}$, $f_E^{in}$, $f_I^{in}$} \label{alg:d}
        \For{$Q \in \{E, I\}$}
            \State $\underline{\nu_Q} \gets F_Q(\underline{\hat{y}})$ \Comment{computed using $A_Q$ and equation \ref{eqn:linear_sys}}
            \State $f_Q^{out} \gets G(f_Q^{in}, \underline{\nu_Q})$ \Comment{computed using equation \ref{eqn:out_rates}}
      \EndFor
      \Return $f_E^{out}, f_I^{out}$

\EndFunction
\end{algorithmic}
\end{algorithm}

\subsection{Supervised Models}
We experiment with four different common supervised learning models for time series regression, \ie convolution neural network (CNN), deep neural network (DNN) (\cite{review_tc}), support vector machine (SVM) (\cite{svm}), and random forest (RF) (\cite{rf}). These models were trained on 99,000 pairs of spike series and parameter labels, and tested on 1,000 other pairs.

The CNN has three layers of convolution followed by 4 fully-connected layers. The DNN consists of four fully-connected layers. They were designed to have roughly the same number of trainable parameters (approximately 30.6M). The details on the architectures, hyper-parameters, and other training configurations are given in the Appendix \ref{sec:model_configs}.

\section{Results} \label{sec:res}
\subsection{Test Performance} \label{sec:test_res}

\begin{figure}
    \hspace*{-1.5cm}
   \includegraphics[width=1.3\linewidth]{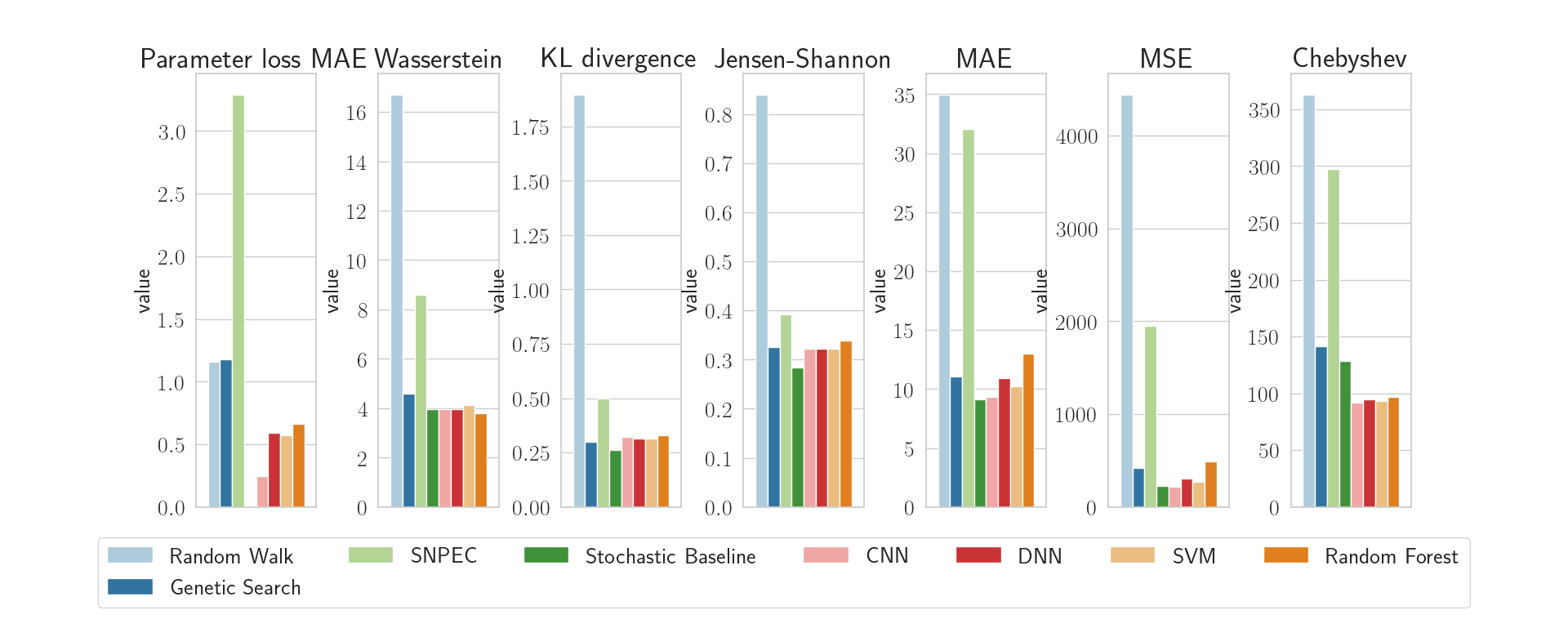}
\caption{The parameter prediction and reconstruction errors of 7 models and a stochastic baseline on 1,000 test examples (lower error is better). The supervised models consistently outperform the classic methods. CNN has the lowest parameter prediction error and comparable reconstruction error to that of the stochastic baseline.}
\label{fig:results}
\end{figure}

We report the performance of four supervised models and three classical models in Figure \ref{fig:results}. The parameter prediction loss is the mean absolute error (MAE) between the predicted and the target parameters. We include six different measures for reconstruction error. Two popular measures of similarity between two spike trains are Victor-Purpura \cite{victor_single} and van Rossum \cite{rossum_single} distances, and their population-level variants \cite{victor, rossum}. However, the
time and space complexity of these distances' computation is polynomial in the number of spikes. These distances have only been used on populations of a handful of neurons, and are too computationally expensive for neuron populations of our size. Instead, we use three measures of information-theoretic distance between two probability distributions, including the Wasserstein distance \cite{wasserstein}, Kullback-Leibler (KL) divergence \cite{kl_div}, and Jensen-Shannon distance \cite{jensen_shannon}. To compute these
distances, we first normalize the spike series of each Q-subpopulation to a probability distribution. Note that these measures are temporal rather than  rate distances. That is, two neuron populations of vastly different firing rates (\ie one population has a much larger higher rate than the other) can have a low information-theoretic distance as long as the times of firing between the two populations tend to coincide. We include three other metrics, \ie mean absolute
error (MAE), mean squared error (MSE), and Chebyshev distance \cite{chebyshev}, which take differences in firing rates into account. 

Because the data-generating neuron model used in this paper is stochastic, two spike series generated from two populations of the exact same parameter set will still almost surely have a non-zero reconstruction error. Thus, we include a ``stochastic baseline" in Figure \ref{fig:results} to capture this. The reconstruction error of that baseline is computed precisely by generating pairs of spike series from the same parameter sets but using different random seeds in the simulation.

The results in Figure \ref{fig:results} show that supervised models are generally far better than the classical methods across metrics. CNN is the best among the supervised learners, achieving low reconstruction error comparable to that of the stochastic baseline. Convolution operators can allow the CNN to encode the temporal correlation in firing events in the spike series better than the DNN and lead to superior performance although CNN and DNN have roughly the same number of trainable network parameters. The random walk tends to be the worst performer since it is the sole mean-field model, taking into account only summary statistics including firing rates and interspike variances, while all other models make full use of spike series. The computation costs of these models are also given in Table \ref{tab:model_times}. CNN is the second computationally most efficient model, only rivaled by DNN. The deep learning models (CNN and DNN) are several order-of-magnitude faster than the
rest \ie 200x faster than random forest, and 8x faster than SVM. Since the supervised models require no additional simulations (each simulation takes around 4-6 seconds with optimized C++ code) or heavy optimization and search at inference time, they are also about 22,000x faster than genetic search, 670,219x faster than SNPE, and 190x faster than random walk. The random walk is the least computationally expensive among the classical methods because although it does require inference-time optimization over the parameter space, it uses an approximate model to compute the reconstruction error instead of running additional simulations. Note that Table \ref{tab:model_times} does not take into account the time required to generate the training dataset for supervised models. However, training data generation is only needed to be done once and the data can be then shared among all supervised model. The data generation time of 99,000 training populations is about 4.5 hours using 24 Haswell CPUs, which is roughly the same time required for testing 8 populations using SNPE or 240 populations using genetic search under the same amount of compute resources.

\begin{table}[ht]
\centering
\caption{Computation cost of models. DNN and CNN were run on a single NVIDIA Tesla P4 GPU. The rest of the models were run parallel on 24 Intel Haswell CPUs. Note that the three classical methods do not have a training stage.}
\begin{tabular}{|c|c|c|}
    \hline 
    \rowcolor[gray]{.9} Models & Training time (seconds/trial/machine) & Inference time (seconds/trial/machine)\\
\hline
CNN & 0.073 & 1.84e-2 \\
\hline
DNN & 0.029 & 7.75e-4 \\
\hline
Random forest & 17.28 & 8.05e-3 \\
\hline
SVM & 0.607 & 2.87e-4 \\
\hline
Genetic search & N/A & 1628 \\
\hline
SNPE & N/A & 48926 \\
\hline
Random walk & N/A & 14.2 \\
\hline
\end{tabular}
\label{tab:model_times}
\end{table}

\begin{figure}
    \hspace*{-2cm}
   \includegraphics[width=1.2\linewidth]{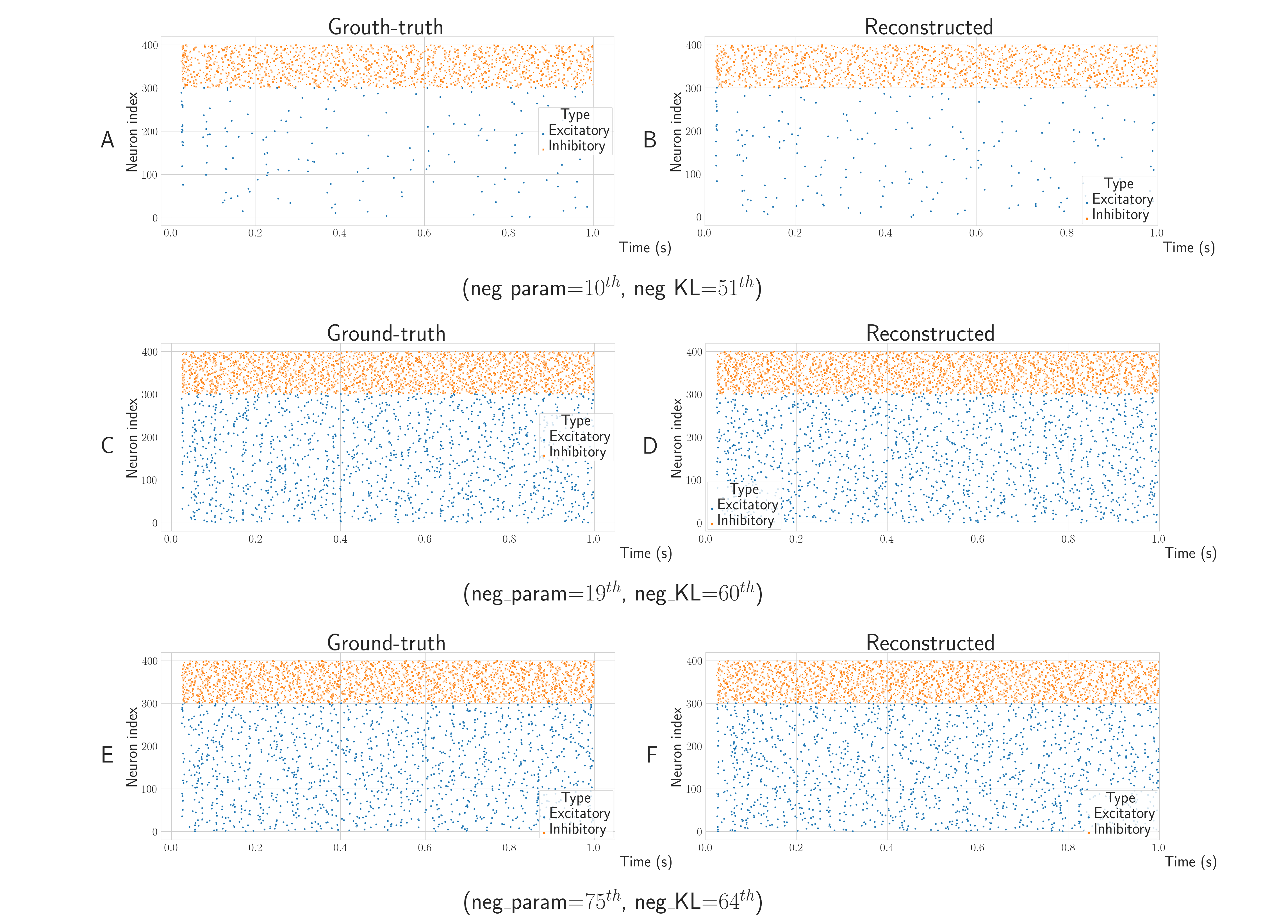}
\caption{A comparison between the ground-truth and reconstructed raster plots using predicted parameter sets from CNN. The ground-truth raster plots are from the test set, and the reconstructed plots are obtained by running the neuronal model simulator with the predicted parameter sets. We show the plots for some examples in the $10^{th}, 19^{th}$ and $75^{th}$ percentile of the negative parameter prediction error of CNN (higher percentile means smaller prediction error). The $75^{th}$ percentile
parameter set yields $64^{th}$ percentile of the negative KL divergence reconstruction error (higher percentile means smaller error). Even, low $10^{th}$ and $19^{th}$ parameter percentiles yield $51^{th}$ and $60^{th}$ percentiles on negative reconstruction error respectively. By comparison, the stochastic baseline averages $63^{th}$ percentile of the negative KL divergence error of CNN.}
\label{fig:reconstructed_raster_plot}
\end{figure}

Figure \ref{fig:reconstructed_raster_plot} compares the ground-truth and reconstructed raster plots given by parameter prediction of the CNN. All of the reconstructed raster plots look visually similar to the ground truth, and are able to capture trends such as degree of synchrony and firing rate. Even predicted parameter sets that have high prediction error ($10^{th}$ and $19^{th}$ in the Figure) can yield reasonably low reconstruction error compared to the stochastic baseline. This is probably because the MFE dynamics has low dimensional feature \cite{cai2022model}. Hence the mapping from the parameter space to the MFE dynamics may be like a projection to the vicinity of a certain unknown low dimensional manifold. As a result, the MFE dynamics is not sensitively dependent on the change of parameters in a certain direction in the parameter space. 

\subsection{Generalization Performance} \label{sec:gen_res}

\begin{figure}
    \hspace*{-2cm}
   \includegraphics[width=1.2\linewidth]{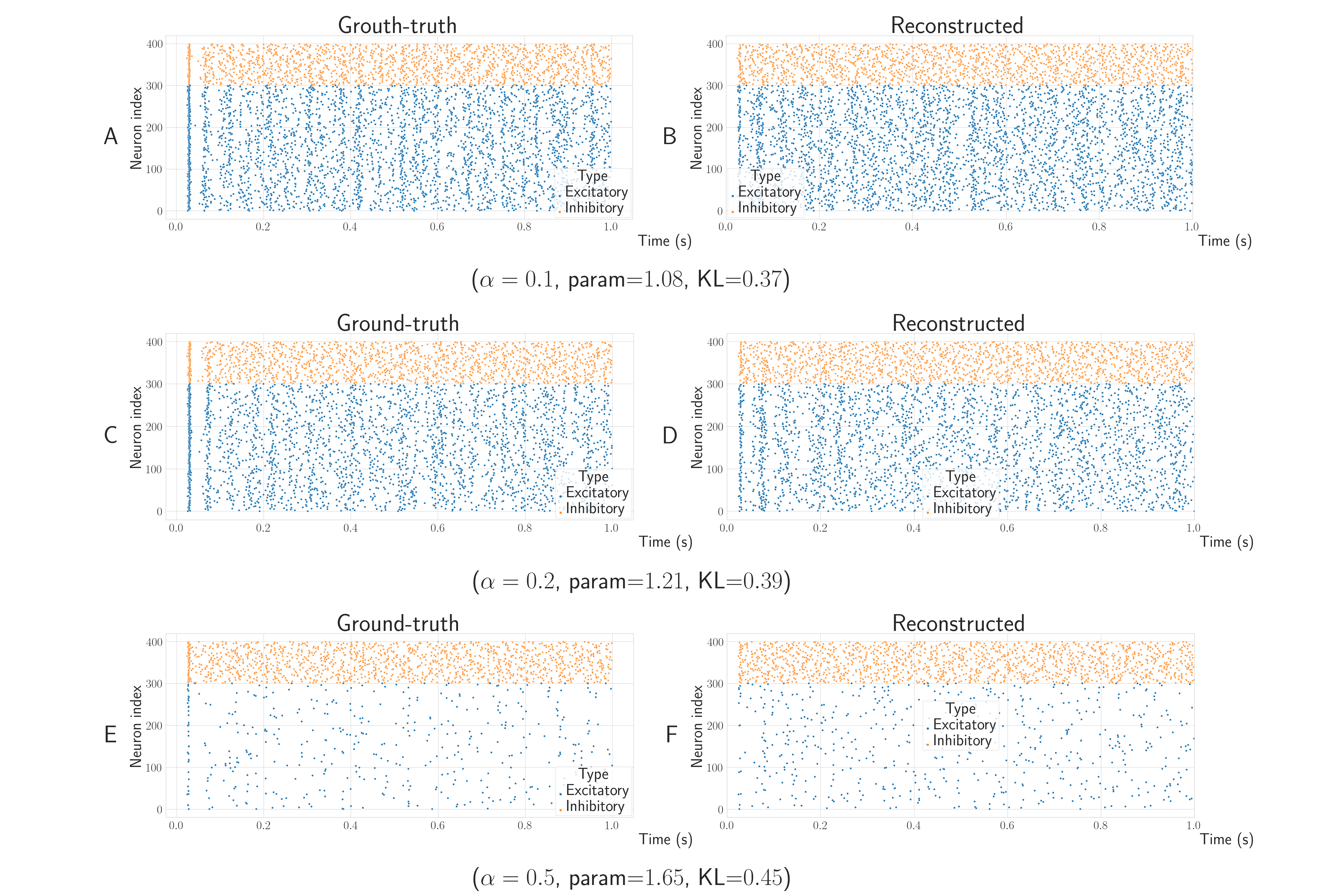}
\caption{A comparison between the ground-truth and reconstructed raster plots using predicted parameter sets from CNN in the generalization experiments. Each of the predicted parameter sets here is the median in parameter prediction error for their corresponding $\alpha$. The parameter prediction error and KL reconstruction divergence are also provided (higher error is worse). By comparison, the stochastic baseline averages $0.262$ in KL divergence. All of the predicted raster plots can capture general patterns such as synchrony in the firing of the ground truths although both of parameter prediction and reconstruction deteriorate with increasing $\alpha$.}

\label{fig:recon_gen}
\end{figure}

\begin{figure}
    \hspace*{-2.75cm}
   \includegraphics[width=1.4\linewidth]{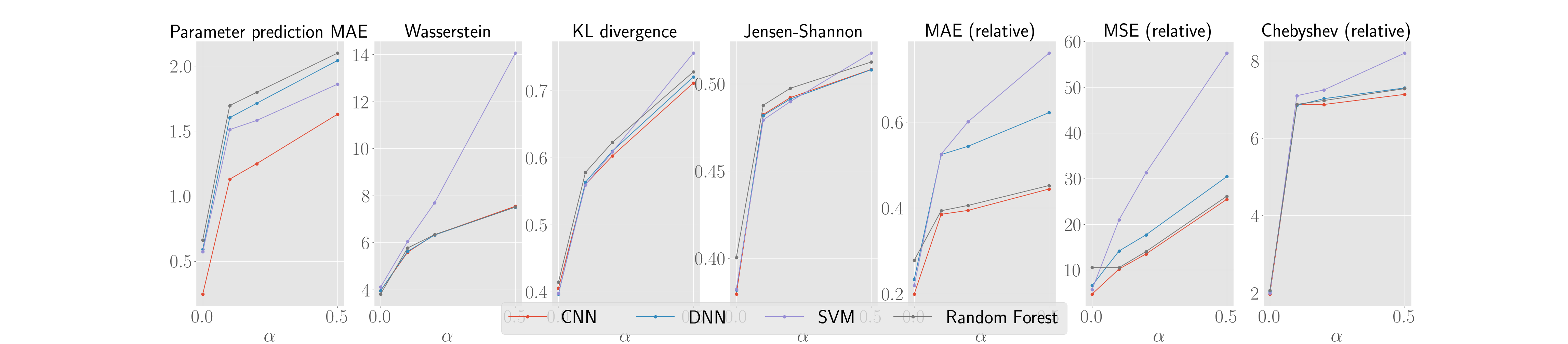}
\caption{Generalization ability of supervised models as a function of $\alpha$ (lower error is better). The parameter $\alpha$ controls the size of the out-of-distribution parameter space, and hence the difficulty of the generalization task. The results on in-distribution test data are also plotted under $\alpha=0$. We normalize the MAE, MSE and Chebyshev reconstruction errors by the average firing rates to account for the difference in firing rates between in-distribution and out-of-distribution data (about 46.7 compared to 14 spikes/neuron/millisecond). To compute information-theoretic distances, spike series were normalized to probability distributions so no further normalization is necessary. CNN tends to generalize the best across metrics.}
\label{fig:generalize}
\end{figure}

We test the supervised models' ability to generalize to out-of-distribution data. We generate the out-of-distribution data as follows. Let $\pi \in \{S_{EE}, S_{IE}, S_{EI}, S_{II}, \tau_E, \tau_I\}$ be any of the varied parameters. Let $[\pi_L, \pi_U]$ be its corresponding range over which an uniform sample is drawn for training and testing in Section \ref{sec:test_res} (see Table \ref{tab:dynamic_params}). Let $D_{\pi} = \pi_U - \pi_L$. We expand $[\pi_L, \pi_U]$ into a wider range $[\tilde{\pi}_L, \tilde{\pi}_U]$ where the length of the new range is increased by an expansion percent $\alpha$ \ie $\tilde{\pi}_U - \tilde{\pi}_L = (1 + \alpha)D_{\pi}$. 
We then sample new test populations uniformly from the range $R_{\tilde{\pi}} = [\tilde{\pi}_L, \pi_L] \cup [\pi_U, \tilde{\pi}_U]$. This procedure generates out-of-distribution testing data such that none of the generated parameters would come from a range already covered by the training data.

For non-negative parameters $\{\tau_E, \tau_I, S_{EE}, S_{IE}\}$, we have
\begin{equation}
    \begin{cases}
        \hat{\pi}_L = \min(\pi_L - \alpha/2 D_\pi, 0) \\
        \hat{\pi}_U = \pi_L + (1+\alpha)D.
    \end{cases}
\end{equation}
For non-positive parameters $\{S_{EI}, S_{II}\}$, 
\begin{equation}
    \begin{cases}
        \hat{\pi}_U = \max(\pi_U + \alpha/2 D_\pi, 0) \\
        \hat{\pi}_L = \pi_U - (1+\alpha)D.
    \end{cases}    
\end{equation}

For each $\alpha \in \{0.1, 0.2, 0.5\}$, we generate $1000$ test
samples and plot the result in Figure \ref{fig:generalize}. For continuity, we also re-plot the test metrics from Figure \ref{fig:results} under $\alpha=0$ (\ie the in-distribution data)\footnote{We are calling in-distribution data as $\alpha=0$ for the purpose for plotting only. Setting $\alpha=0$ and follows the generalization data generation procedure will actually produce an empty parameter space.}. We see that parameter prediction and reconstruction error increase with the difficulty of the generalization task $\alpha$. We also compare the raster plots of the predicted and ground truth parameters in these generalization domains in Figure \ref{fig:recon_gen}. Similar to Figure \ref{fig:reconstructed_raster_plot}, all reconstructed raster plots are visually similar to the ground truth, despite having much bigger error in the parameter space. This result further support our conjecture that the spiking pattern is not sensitively dependent on the change of parameters in some directions. 

Classical methods do not experience this out-of-distribution degradation since they produce a predicted parameter set on a case-by-case basis \ie a new search is initiated whenever a new target spike series is presented. In contrast, supervised models maintain some internal parameters (\eg the weights of the CNN) that can carry learned knowledge from one example to the next. Thus, testing for generalization of these models is necessary.

\section{Conclusion and Future Work} \label{sec:con}

In this paper, we formulate the parameter estimation task as a supervised learning problem. A variety of supervised models were trained, and compared against some representative classical methods. The result is promising. The supervised models can learn the parameter set of a neuronal network from its firing dynamics, and reconstruct the spiking pattern reasonably. They can also have some generalization ability to out-of-distribution data. This tells us that although the spike count in a multiple firing event (MFE) has very high volatility due to the randomness in the modeling (\cite{li2019stochastic}), its spiking pattern is fairly robust. We can use machine learning to find an almost deterministic relation between parameter sets and spiking patterns. This gives us confidence that the dynamics in MFEs can be described by a lower dimensional dynamical system. We remark that the low dimensional characteristics of MFE dynamics is corroborated by the recent model reduction study in \cite{cai2022model}. We will further investigate this model reduction problem in our future studies. 

This paper also sets up an example of parameter tuning for biological models at a population level with high dimensional random dynamics. When the output of a model lives in a high dimension space and the dynamics of this model is very
complicated, the dependency of model output and parameter can be highly nonlinear. It is not easy to use traditional statistical methods or optimization methods to guide the parameter tuning. In fact, the two-stage optimization method using a random walk approximate model that we experimented with fails to produce desired performance. Modern machine
learning could help us on this front. We can generate a large set of parameter-output pairs, and send them into a supervised learner that can approximate high dimensional nonlinear functions. This ``teaches'' the supervised model the relation between parameter sets and neuronal model outputs. Then the supervised learner can help us to find a suitable parameter corresponding to the desired output. We expect this to be extended to a wider range of problems in mathematical biology. These parametric supervised models are also several order-out-magnitude faster than the classical methods that rely on additional simulation at inference time on a test-case-by-test-case basis. Since the neuronal network model studied in this paper can produce fairly diverse MFE dynamics, another future research direction is to use supervised learning approaches to calibrate the neuronal data-generating model to real biological data such as from Electroencephalography ( \cite{MCMC}) or to synthetic data produced by a different neuronal model.

\section*{Conflict of Interest Statement}

The authors declare that the research was conducted in the absence of any commercial or financial relationships that could be construed as a potential conflict of interest.

\section*{Author Contributions}
YL conceived the original ideas of the study. LL and YL wrote the simulation code. LL performed model training, experiments and analysis. LL wrote the first draft of the manuscript. Both authors contributed to manuscript revision, read, and approved the submitted version.

\section*{Funding}
YL is partially supported by NSF DMS-1813246 and NSF DMS-2108628. LL was supported by the University of Massachusetts' Sheila Flynn research scholarship. 

\section*{Acknowledgments}
The authors would like to acknowledge the helpful discussion with Professors Yaoyu Zhang (Shanghai Jiao Tong University, China) and Lai-Sang Young (New York University).


\section*{Data Availability Statement}
The datasets generated for this study can be found in the Google Folder \url{https://drive.google.com/drive/folders/1ae0ZyfdhbEE8XvBvHabTMOxIhre-eyfc?usp=sharing}.

\section*{Appendices}
\appendix
\section{Parameter settings} \label{sec:pars}
The fixed values of 11 static parameters are given in Table \ref{tab:static_params}. 
\begin{table}[ht]
\centering
\caption{Static parameters.}
\begin{tabular}{|c|c|}
\hline 
\rowcolor[gray]{.9} Parameter & Value \\
\hline
$M$                              & 100 \\
    \hline
$-M_r$                           & -66 \\
     \hline
$\lambda_E, \lambda_I$           & 3000, 3000 (spikes/sec) \\
     \hline
$N_E, N_I$                       & 300, 100  (neurons)\\
     \hline
$P_{EE}, P_{IE}, P_{EI}, P_{II}$ & 0.15, 0.5, 0.5, 0.4 \\
     \hline
$\tau_R$                         & 2.5 (ms) \\
     \hline
\end{tabular}
\label{tab:static_params}
\end{table}

Each of the 6 dynamic parameters is sampled uniformly from some pre-specified range to generate a training dataset. The training ranges are given in Table \ref{tab:dynamic_params}.

\begin{table}[ht]
\centering
\caption{The ranges of dynamic parameters}
\begin{tabular}{|c|c|}
    \hline 
\rowcolor[gray]{.9} Parameter & Range \\
\hline
$S_{EE}$ & $[3.0, 7.0]$ \\
\hline
$S_{IE}$ & $[1.0, 5.0]$ \\
\hline
$S_{EI}$ & $[-4.0, -0.5]$ \\
\hline
$S_{II}$ & $[-4.0, 0.0]$\\
\hline
$\tau_E$ & $[0.5, 2.5]$\\
\hline
$\tau_I$ & $[2.0, 6.0]$ \\
\hline
\end{tabular}
\label{tab:dynamic_params}

\end{table}

\section{Model Configurations} \label{sec:model_configs}
\subsection{Genetic Search}
The genetic search we used has 8 chromosomes within a generation and runs for 20 generations. In each generation, there are 4 mating parents selected using steady-state selection. We have single point cross-over and adaptive mutation \cite{adaptive_mutation}, which uses fitness values in computing mutation probabilities and favors high-quality solutions. The mutation probabilities for high and low-quality solutions are $0.25$ and $0.5$ respectively. The search stops if the
fitness saturates for 7 consecutive generations. The initial population of examples is sampled from the data-generating distribution in Table \ref{tab:dynamic_params}.

\subsection{SNPE}
SNPE starts out with 1,000 pilot examples drawn from the data-generating distribution in Table \ref{tab:dynamic_params}. SNPE then runs for 4 rounds, sampling 500 examples each round drawn from the current estimate of the parameter posterior distribution. The conditional density estimator is a deep neural network with the exact same architecture as the DNN in the supervised approach (see \ref{sec:dnn_arch}) with masked autoregressive flow \cite{maf} and 5 MADES \cite{made}. At each
round, the density estimator neural network is trained for 50 epochs.

\subsection{Random Forest}
Our random Forest consists of 1,000 trees, each with a maximum depth of 20. To regularize the training, the maximum number of features to consider when looking for the best split is limited at $\log_2$  of the number of features.

\subsection{Support Vector Machine}
We fit one support vector regression model per each of the 6 target parameters. The SVM uses a radial basis function kernel \cite{rbf} with $\gamma = 1/(\textrm{no. of features} \cdot Var(\textrm{input}))$, and regularization parameter $C=1$ and $\epsilon=0.1$.

\subsection{DNN} \label{sec:dnn_arch}
The architecture of the DNN is given below.
\begin{verbatim}
DNN(
  (fc1): Linear(in_features=800, out_features=2000, bias=True)
  ReLU()
  (fc2): Linear(in_features=2000, out_features=2010, bias=True)
   ReLU()
  (fc3): Linear(in_features=2010, out_features=4096, bias=True)
   ReLU()
  (fc4): Linear(in_features=4096, out_features=4096, bias=True)
   ReLU()
  (out): Linear(in_features=4096, out_features=6, bias=True)
)
\end{verbatim}

\subsection{CNN}
The architecture of the CNN is given below.
\begin{verbatim}
 CNN(
  (convLayer1): Sequential(
    (0): Conv1d(1, 150, kernel_size=(21,), stride=(1,))
    (1): BatchNorm1d(150, eps=1e-05, momentum=0.1)
    (2): ReLU()
    (3): MaxPool1d(kernel_size=5, stride=5, padding=0, dilation=1)
  )
  (convLayer2): Sequential(
    (0): Conv1d(150, 250, kernel_size=(21,), stride=(1,))
    (1): BatchNorm1d(250, eps=1e-05, momentum=0.1)
    (2): ReLU()
    (3): MaxPool1d(kernel_size=4, stride=4, padding=0, dilation=1)
  )
  (convLayer3): Sequential(
    (0): Conv1d(250, 500, kernel_size=(21,), stride=(1,))
    (1): BatchNorm1d(500, eps=1e-05, momentum=0.1)
    (2): ReLU()
    (3): MaxPool1d(kernel_size=7, stride=7, padding=0, dilation=1)
  )
  (fc1): Linear(in_features=1000, out_features=2048, bias=True)
  (fc2): Linear(in_features=2048, out_features=4096, bias=True)
  (fc3): Linear(in_features=4096, out_features=4096, bias=True)
  (out): Linear(in_features=4096, out_features=6, bias=True)
)
\end{verbatim}

Both the DNN and CNN use Adam optimizer with a learning rate = 5.4e-05. The learning rate was found by random hyper-parameter search on a log-scale grid of learning rates from $10^{-6}$ to $10^{-1}$ using a small dataset of 5,000 spike trains, and 5-fold cross-validation.

\bibliography{pub}


\end{document}